\documentstyle[12pt]{article}
\def\tc{\tilde c}

\def\tp{\tilde \psi}
\def\op{\overline \psi}
\def\top{\tilde {\overline \psi}}
\def\tP{\tilde \Psi}
\def\oP{\overline \Psi}
\def\toP{\tilde {\overline \Psi}}
\def\t{\theta}
\def\ss{( x,\theta )}
\def\no{\nonumber}
\begin{document}
\begin{center}
{\large {\bf Superspace Formulation for the BRST Quantization of the
Chiral Schwinger Model}}\\
\end{center}
\vspace{1cm}

{\large Everton M. C. Abreu $^{1,2}$ and   Nelson R. F. Braga $^{1}$} \\
\vspace{1cm}

$^1$ Instituto de F\'\i sica, Universidade Federal  do Rio de Janeiro,\\
Caixa Postal 68528, 21945  Rio de Janeiro,
RJ, Brazil\\

$^2$ Instituto de F\'\i sica, Universidade do Estado do Rio de
Janeiro,\\ Rua Sao Francisco Xavier 524,CEP 20550 Rio de Janeiro, RJ,
Brazil
\vspace{0.5cm}
\abstract
It has recently been shown that the Field Antifield quantization
of anomalous irreducible gauge theories with closed algebra can be represented in a BRST superspace where the quantum action at one loop order, including the Wess  Zumino term, and the anomalies show up as components of the same superfield.
We show here how the Chiral Schwinger model can be represented in this formulation.

\vskip3cm
\noindent PACS: 11.15 , 03.70
\vfill\eject
\section{Introduction}
The Lagrangian BRST formalism of Batalin and Vilkovisky (BV) \cite{BV,HT,GPS} is presently considered the most powerful
procedure for the quantisation of gauge theories.  The application 
of this formalism to anomalous gauge theories was first discussed by
Troost, van Nieuwenhuizen and Van Proeyen \cite{TPN}, that succeeded in using the Pauli Villars regularization  in order to give a regularized meaning to the master equation at one loop order.
The possibility of more general regularizations that enable the quantization of more complex gauge theories where the higher loop order terms play an important role  is presently under study\cite{Paris,DPT}.
For the case of irreducible gauge theories with closed algebra,
it was shown that the BV procedure can be formulated in a superspace with one Grassmanian variable\cite{BD,AB}, where
BRST transformations are realised as translations.

The Chiral Schwinger Model (CSM) has been an important device to 
 understand the quantisation of anomalous gauge theories.
It was shown that the Wess Zumino term that restores the gauge 
invariance at the quantum level, realizing the mechanism proposed by
Faddeev and Shatashvilli\cite{FS} can be generated by just solving 
the (regularized) master equation at one loop order\cite{BM} for this model, reproducing the results that have been found before by Harada and Tsutsui\cite{Harada} in a tricky way.

The superspace formulation  has the nice property of been, by construction, explicitly BRST invariant. The master equation is 
translated into the existence of a superfield structure associated to the quantum action in such a way that realizing the Wess Zumino
mechanism is just equivalent to building up such a superfield without anomaly.
We will see that the Chiral Schwinger Model illustrates this 
superspace formulation in an interesting way. 
This article is organized as follows: in section {\bf 2} we briefly
review the superspace formulation for the BV formalism at one loop
order. In section {\bf 3} we present our results for the CSM and 
section {\bf 4} is devoted to some concluding remarks.

\section{Superspace Formulation for the BV formalism}
The fermionic nature and the nillpotency of the BRST transformations makes it possible to build up a superspace representation  where they are realized as
translations in a Grassman variable\cite{SUP}.
One adds to the original space-time
variables one Grassmanian degree of freedom $\theta$ 
and associate to each original field $\phi (x)$ a
superfield:

\begin{equation}
\label{SF}
\Phi (x , \theta ) = \phi (x) + \theta \delta \phi (x)
\end{equation}

\noindent in such a way that:

\begin{equation}
\delta_{_{BRST}}\Phi \ss = {\partial  \over \partial\theta }\,\Phi\ss
\end{equation}

When one tries to apply this superspace realization to the case of the field antifield (FA) quantisation one faces a problem.
The quantum master equation\cite{BV}

\begin{equation}
\label{Master}
{1\over 2}(W,W) = i\hbar\Delta W
\end{equation}

\noindent involves the operator:

\begin{equation}
\label{Delta}
\Delta \equiv
{\delta_r \over\delta \phi^a }{ \delta_l \over \delta \phi^\ast_a }\;
\end{equation}

\noindent that represents the possibly non trivial behavior of the
path integral measure.One then  needs a superspace version for this operator and thus one should introduce functional derivatives with respect to superfields. However, superfields of the form (\ref{SF}) will in general be constrained, as the BRST transformations are in general not independent of the corresponding fields.
For an unconstrained superfield:

\begin{equation}
\label{SFG}
\Omega (x , \theta ) = A (x) + \theta B (x)
\end{equation}

\noindent where 

\begin{equation}
\label{INDEP}
{\delta B(x)\over \delta A(x^{\prime})} \,=\,0\,\,\,\,;
\,\,\,\,\,\,
{\delta A(x)\over \delta B(x^{\prime})}\,=\,0
\end{equation} 

\noindent we can define a functional  derivative and relate it to the derivatives with respect to the component fields:

\begin{equation}
\label{COMP}
{\delta \over 
\delta \Omega (x ,\theta )} = 
\,{\delta \over 
\delta B (x )} 
+\,\theta {\delta \over 
\delta A (x )} 
\end{equation}

However, as discussed in \cite{AB}, functional differentiation and integration for constrained superfields  is not in general well defined.
This problem was circumvented in \cite{AB} by applying a procedure due to Alfaro and Damgaard\cite{COL}\footnote{See also \cite{DJ} for a review}, that we will call collective field approach to BV, where one gets the BV action by trivially enlarging the field content of the theory and then choosing an appropriate gauge fixing structure. Following this approach, as we will see, one is able to build up unconstrained superfields even when the original
BRST transformations are not independent of the associated fields. 

The collective field approach to BV consists (in a very summarized way) in
starting with  a gauge field theory characterized by a classical action
$S_0[\phi^i] $, introducing ghosts, antighosts and auxiliary fields
associated to the original gauge invariances of  $S_0$ in the usual way, getting an enlarged field set represented as $\phi^A$.  These fields realize a BRST
algebra:

\begin{equation}
\label{OGS}
\delta_0 \phi^A = R^A \,[\phi ]
\end{equation}

Then we introduce a new set of fields called collective fields 
$\tilde \phi^A$ and replace everywhere $\phi^A$ by 
$\phi^A - \tilde \phi^A$. 
This way we double the field content of the theory
and at the same time associate to each field a new trivial shift symmetry.
In order to gauge fix these new symmetries we introduce new ghosts, 
antighosts and auxiliary fields, represented respectively as:
$\pi^A$, $\phi^{\ast\,A}$ and $B^A$.
We have a large freedom in choosing the BRST transformations
for this enlarged set of fields. Following \cite{COL} we can define the enlarged BRST algebra as 

\begin{eqnarray}
\label{ALG}
\delta \phi^A &=& \pi^A\nonumber\\
\delta \tilde \phi^A &=& \pi^A - R^A [\phi -\tilde \phi ]
\nonumber\\
\delta \pi^A &=& 0\nonumber\\
\delta \phi^{\ast\,A} &=& B^A\nonumber\\
\delta B^A &=& 0
\end{eqnarray}

\noindent and the total action as

\begin{equation}
\label{A1}
S = S_0 [\phi^i -\tilde \phi^i ] - \delta (\phi^{\ast\,A}
\tilde \phi^A ) + \delta \psi [\phi^A]
\end{equation}

\noindent where $\psi [\phi^A]$ is a fermionic functional representing the gauge fixing of the original symmetries (\ref{OGS}). The BV gauge fixed classical action is obtained if one functionally integrates 
the vacuum functional associated with $S$ over $\pi^A$, $\tilde \phi^A$ and $B^A$.

The interesting point is that in this collective field approach all the  fields of the sets $\phi^A$ and $\phi^{\ast\,A}$ have BRST transformations that are 
independent quantities, unrelated to the associated field, as follows from (\ref{ALG}).
Therefore, if we introduce superfields of the form (\ref{SF}),
at least for this two sets, they will be unconstrained.
The component decomposition (\ref{COMP}) for the functional derivatives then makes it easy to see that the operator:

\begin{equation}
\label{DS}
\underline \Delta \equiv 
\int dx  \int d\theta \int d\theta^\prime\,{\delta_r \over 
\delta \Phi^A (x,\theta ) }\,\,
{\delta_l \over 
\delta \Phi^{\ast\,A} (x ,\theta^\prime \,)}
\end{equation}

\noindent with

\begin{eqnarray}
\label{EXTALG}
\Phi^A (x,\theta ) &=& \phi^A (x) + \theta \pi^A (x) \nonumber\\
\tilde \Phi^A (x,\theta ) &=& \tilde \phi^A (x) + \theta 
( \pi^A (x) - R^A [\,\phi -\tilde \phi \,]\, )\nonumber\\
\Phi^{\ast\,A} (x,\theta ) &=& \phi^{\ast\,A} (x) + \theta B^A (x) \nonumber\\
\end{eqnarray}

\noindent represents the operator $\Delta$ in superspace.

The naive action of this operator leads to an undefined 
singular result because of the two functional derivatives acting in the same space time point.
It was shown in \cite{AB} that the Pauli Villars regularization developed in \cite{TPN} that gives a meaning 
to the one loop order master equation can be applied in the superspace formulation. We will see how to do it in next section.

In the superspace formulation the quantum action, for non anomalous gauge theories, will have the component expansion

\begin{equation}
\label{SA}
\underline W = W + \theta i\hbar \Delta W 
\end{equation}
 
\noindent and the master equation will read:

\begin{equation}
\label{SMaster}
{\partial \over \partial \theta}  \underline W \,=\, 
i\hbar \underline \,\,\underline\Delta \underline W 
\end{equation}

\noindent corresponding,  order by order in $\hbar$, (we are considering 
quantum corrections only up to one loop order):

\begin{equation}
\label{SMO}
{\partial \over \partial \theta}  \underline S \,=\, 
0 \,\,\,\,;\,\,\,\,
{\partial \over \partial \theta}  \underline M_1 \,=\, 
i \underline \Delta \,\, \underline S 
\end{equation}

\vskip 1cm 

\section{The Chiral Schwinger Model}
The classical action for the Chiral Schwinger Model (CSM) is:

\begin{equation}
S_0 = \int d^2x \Big[ -{1\over 4} F_{\mu\nu} F^{\mu\nu} +
{1\over 2}\overline \psi {\slash\!\!\!\!D} (1-\gamma_5) \psi \Big]
\end{equation}

\noindent where $D_\mu = \partial_\mu + ie A_\mu$. The BRST transformations of the fields are:

\begin{eqnarray}
\label{TS}
\delta A_\mu &=& \partial_\mu c\nonumber\\
\delta \psi &=& i\psi c\nonumber\\
\delta \overline\psi &=& - i \overline \psi c
\nonumber\\
\delta c &=& 0
\end{eqnarray} 

\noindent where $c$ is the ghost field associated to the gauge invariance of $S_0$.

We enlarge, as explained in the previous section, the field 
content of the theory, introducing the collective fields 
$\tilde A_\mu , \tilde \psi , \tilde {\overline \psi} ,
\tilde c $ and build up the associated superfields:

\begin{eqnarray}
\underline A_\mu (x,\theta) &=& A_\mu (x) + \theta 
\pi ^{\,[\,A_\mu\,]}_\mu (x)\nonumber\\
\underline {\tilde A}_\mu  (x,\theta) &=& \tilde A_\mu (x) 
+ \,\theta\, \Big[ \,\pi ^{\,[\,A_\mu\,]}_\mu (x) 
\,-\,
\partial_\mu ( c(x) - \tilde c(x))\,\Big]
\nonumber\\
\underline A_\mu^\ast (x,\theta) &=& A_\mu (x) + \theta 
B_\mu^{\,[\,A_\mu\,]}(x)\nonumber\\
& &\no\\
\Psi \ss &=& \psi( x ) + \t \pi^{\,[\,\psi\,]} (x)\nonumber\\
\tP \ss &=& \tp (x) + \t \,\big[\, \pi^{\,[\,\psi\,]} (x) - i (\psi (x) - \tp (x) )
( c(x) - \tc (x) ) \Big] \no\\
\Psi^\ast \ss &=& \psi^\ast (x) + \t B^{\,[\psi\,]} (x) \no\\
& &\no\\
\oP \ss &=& \op ( x ) + \t \pi^{\,[\,\op\,]} (x)\nonumber\\
\toP \ss &=& \top (x) + \t \, \big[ \, \pi^{\,[\,\op\,]} (x) + i (\op (x) - \top (x) )
( c(x) - \tc (x) ) \Big] \no\\
\overline \Psi^\ast \ss &=& \op^\ast (x) + \t B^{\,[\op\,]} (x) \no\\
& &\no\\
\eta \ss &=& c (x) + \t \pi ^{\,[\,c\,]} (x) \no\\
\tilde\eta \ss &=& \tilde c (x) + \t  \pi ^{\,[\,c\,]} (x)\no\\
\eta^\ast \ss &=& c^\ast (x) + \t B^{\,[\,c\,]} (x) 
\end{eqnarray}

The total superfield action will be:

\begin{equation}
\underline S = \underline S_0 + \underline S_1 + \underline S_2
\end{equation}

\noindent with the extended superspace version of the classical
action:

\begin{eqnarray}
\label{S0}
\underline S_0 &=& \int d^2x \Big( -{1\over 4} F_{\mu\nu}\,
\big[ \underline A_\mu\,-\,\underline {\tilde A}_\mu\, \big]
 F^{\mu\nu} \big[\underline A_\mu\,- \underline {\tilde A}_\mu\,\big] \nonumber\\
 &+& {1\over 2}( \overline \Psi - \toP ){\slash\!\!\!\!D}\,
[\underline A_\mu\,- \underline {\tilde A}_\mu\,]
 (1-\gamma_5) ( \Psi - \tP) \Big)
\end{eqnarray}

\noindent the gauge fixing of the shift symmetry:

\begin{equation}
\underline S_1 = - {\partial \over \partial\t } \int d^2 x 
\Big[ {\underline A}^\ast_\mu \underline {\tilde A}^\mu +
 \Psi^{\ast} \,\tP \,+\, \oP^\ast\, \toP +\, \eta^\ast\, \eta \,\Big]
\end{equation}

\noindent and the gauge fixing of the original symmetry of $S_0$:

\begin{equation}
\underline S_2 = {\partial \over \partial\t } \int d^2x 
\,\Lambda\, \big[ {\underline A}_\mu , \Psi, \oP , \eta \big] 
\end{equation}

\noindent with the collective field version of the classical action:

We must now build up a superspace Pauli Villars (PV) action, that will regularize the action of the operator $\underline \Delta$ on the action. We only need PV partners for the fermionic fields as we can easily see from  (\ref{TS}) that
the gauge field and the ghost will not contribute to $\Delta S$  as their transformations are independent of the field itself. Following the prescriptions of \cite{AB} we
associate  with $\psi$ and $\overline\psi$ the PV fields 
$\chi$ and $\overline\chi$ and the corresponding collective tilde fields and introduce the action:

\begin{eqnarray}
\underline S_{PV} &=&   {1\over 2} (\underline{\overline \chi} - 
\underline {\tilde {\overline \chi}} ) \,{\slash\!\!\!\!D}\,
[\,{\underline A}_\mu - {\tilde {\underline A}}_\mu \,]
\,(\underline \chi - \underline {\tilde \chi} )\nonumber\\
&-&{1\over 2} M (\underline {\overline \chi} - 
\underline {\tilde {\overline \chi}} ) 
(\underline \chi - \underline {\tilde \chi} ) 
- {\partial\over \partial \theta}\,
\Big(\,\underline {\overline \chi}^{\ast\,} \underline {\tilde {\overline \chi}} + 
\,\underline \chi^{\ast\,} \underline {\tilde \chi} \Big)
\end{eqnarray}
 
\noindent that represents a copy of the fermionic part of action (\ref{S0}) but with a mass term that, after calculating the regularized $\delta S$, allows the removal of 
the  PV fields by taking the limit $\,M\,\rightarrow \infty$.

We define the BRST transformations of the PV fields to be similar to the ones from the corresponding fields given by (\ref{TS}).
Thus the PV  superfields will have the structure:

\begin{eqnarray}
\label{PVSUP}
{\underline \chi} (x,\theta ) &=& \chi (x) + \theta \,
\,\pi^{[\,\chi\,]\,}
\nonumber\\
{\tilde {\underline \chi}} (x,\theta ) &=& \tilde \chi (x) + \theta 
\Big( \,
\,\pi^{[\,\chi\,]\,} -  i (\chi - \tilde\chi ) 
( c - \tilde c )  \, \Big)\nonumber\\
{\underline \chi}^{\ast\,} (x,\theta ) &=& \chi^{\ast\,} (x) + 
\theta B^{\,[\,\chi\,]\,}
\nonumber\\
\overline {\underline \chi} (x,\theta ) &=& 
\overline\chi (x) + \theta \,
\,\pi^{[\,\overline\chi\,]\,}
\nonumber\\
{\tilde {\overline {\underline \chi}}} (x,\theta ) 
&=& \tilde { \overline \chi} (x) + \theta 
\Big( \,
\,\pi^{[\,\overline\chi\,]\,} +  i (\overline \chi - \overline {\tilde\chi} ) 
( c - \tilde c )  \, \Big)\nonumber\\
\overline {{\underline \chi}}^{\ast\,} (x,\theta ) &=& \overline\chi^{\ast\,} (x) + 
\theta B^{\,[\,\overline\chi\,]\,}
\end{eqnarray}

\noindent 

As usual, the PV fields are defined formally in such a way that their 
one loop contributions have a minus sign relative to the original fields\cite{TPN}.
The action of the operator $\underline \Delta $  on the 
regularized  total action, if we include the PV fields
also in the operator  
(\ref{DS}), is then:

\begin{equation}
\label{DSR}
\underline \Delta (\underline S + \underline S_{PV} ) 
\, = \, 0
\end{equation}

\noindent 
The regularized form of $\Delta S$ when we use the PV regularization shows up as a violation of the zero order master equation 
associated  to the presence of the mass term.
In the present superspace formulation, this absence of
BRST invariance of the total (regularized) classical action $S_{T} =
S + S_{PV}$ is translated into the presence of a $\theta$ component in
the corresponding superfield:

\begin{equation}
\underline S_{T} = \underline S + \underline S_{PV} = 
S_T + \theta \delta S_T
\end{equation}

The general form of $\delta S_T$ is

\begin{equation}
\label{1}
\delta S_T = i M \, (\overline\chi - \tilde {\overline \chi} ) 
(\chi - \tilde\chi )\,( c - \tilde c )
\end{equation}

\noindent Integration over the fields $\pi^{[\,\chi\,]\,A\,}
\,,\,  B^{\,[\,\chi\,]\,A}$ and $\tilde \chi^A$ removes the 
extended collective field structure, recovering  the
usual result as in \cite{TPN}, that corresponds in (\ref{1})
 just to the absence 
of the collective tilde fields.
The next step would be to integrate over the PV fields. We will
not repeat this procedure here as it is exactly the same as in the component case, that  is widely discussed in the
literature\cite{GPS,TPN,DJ}. The result is:

\begin{equation}
(\Delta S)_{reg.} \,=\, {i\over 4\pi}
\int d^2x ( c -\tilde c) \Big( (1-a) \partial_\mu (A^\mu - \tilde A^\mu ) - \epsilon^{\mu\nu}\partial_\mu 
(A_\nu -\tilde A_\nu ) \Big)
\end{equation}

Now going back to the one loop order master equation (\ref{SMO}) we have to look for a superfield $\overline M_1$ whose
$\theta$ component is equal to $i (\Delta S)_{reg.}$.
It is well known that for this model, that possess a genuine anomaly, a local solution for the master equation can not be found in the original space of fields and antifields.
However the  Wess Zumino mechanism of restoring gauge invariance can be realized in the field antifield 
formalism\cite{BM2,GP1} by introducing a field (and the corresponding antifield) that transforms as the gauge group elements. In the present  superspace formulation  we can introduce  the superfields 

\begin{eqnarray}
\Omega  (x,\theta) &=& \omega (x) + \theta 
\pi^{\,[\,\omega\,]} (x)\nonumber\\
\tilde\Omega (x, \theta) &=& \tilde\omega (x) 
+ \,\theta\, \Big[ \,\pi^{\,[\,\omega\,]} (x) 
\,-\,
( c - \tilde c)\,\Big]
\nonumber\\
\Omega^\ast (x,\theta) &=& \omega^\ast + \theta 
B^{\,[\,\omega\,]}(x)
\end{eqnarray}

\noindent that realize in superspace the collective field version of
the Wess Zumino field transformations. 
We include a gauge fixing term for the WZ field in the action defining:

\begin{equation}
\underline S^\prime = \underline S 
- {\partial\over\partial\theta} \, \int d^2 x 
\, \tilde\Omega \, \Omega^\ast 
\end{equation}

From the transformation of $\Omega$ one easily realizes that
$\Delta S= \Delta S^\prime$. 
Now we can write a superfield that satisfies $ \partial {\overline M}_1 / \partial \theta 
= i\,(\Delta S)_{reg.}$. 

\begin{eqnarray}
{\overline M}_1 &=& {1\over 4\pi} \int d^2x \Big(
{(a-1)\over 2} \partial_\mu (\Omega - \tilde\Omega) 
\partial^\mu (\Omega - \tilde\Omega) \nonumber\\
&-& \partial_\mu (\Omega - \tilde\Omega) 
\big( (a-1) (\underline A^\mu - \underline {\tilde A}^\mu) 
+ \epsilon^{\mu\nu}  ( \underline A_\nu - 
\underline {\tilde A}_\nu ) \big) \Big)
 \end{eqnarray}

\noindent in components this superfield reads:
$\underline M_1 = M_1 + \theta (i\Delta S)_{reg.}$ with:

\begin{eqnarray}
M_1 &=& {1\over 4\pi} \int d^2 x \Big(
{(a-1)\over 2} \partial_\mu (\omega - \tilde\omega) 
\partial^\mu (\omega - \tilde\omega) \nonumber\\
&-& \partial_\mu (\omega - \tilde\omega) 
\big( (a-1) ( A^\mu -  {\tilde A}^\mu) 
+ \epsilon^{\mu\nu}  (  A_\nu - 
{\tilde A}_\nu ) \big) \Big)
\end{eqnarray}

\noindent If we remove the collective fields, this corresponds just to the Wess Zumino term found in \cite{BM}
in the non superspace approach.

Therefore, the superfield $\underline W = 
\underline S^\prime + \underline M_1 $ satisfies the superspace version of the master equation  (\ref{SMaster}),
representing the superfield action, that includes, besides 
the quantum action, also the anomalous contribution from the
path integral measure $\Delta S$.

\section{Conclusion}
We have shown to represent the quantum action of the Chiral Schwinger model in a BRST superspace.
An interesting point of this formulation is that both the
action and the $\Delta S$ term (that comes from the non trivial behaviour of the path integral measure) show up in the same superfield. The master equation corresponds thus just to a restriction on the structure of this object.
We have also shown that the Wess Zumino mechanism can also be 
realised in this formulation, by adding a superfield that represents  the gauge group elements.

\vskip 1cm
\section{Acknowledgements}This work was partially  supported  by CNPq, FINEP, 
FUJB and CAPES
(Brazilian Research Agencies).
\vfill\eject


\begin{thebibliography}{30}
\bibitem{BV} I. A. Batalin and G. A. Vilkovisky, Phys. Lett. B102
(1981) 27, Phys. Rev. D28 (1983) 2567.
\bibitem{HT}M.Henneaux and C.Teitelboim, Quantization of Gauge Systems,
Princeton University Press 1992, Princeton, New Jersey.
\bibitem{GPS} J. Gomis, J. Paris and S. Samuel, 
Phys.  Rep. 259 (1995) 1.
\bibitem{TPN}  W.Troost, P.van Nieuwenhuizen and
A. Van Proeyen, Nucl. Phys. B333 (1990) 727.
\bibitem{Paris} J.Paris, Nucl. Phys. B450 (1995) 357.
\bibitem{DPT} F.De Jonghe, J.Paris and W.Troost, "The BPHZ renormalised BV master equation and Two loop Anomalies in Chiral Gravities" HEP-TH 9603012.
\bibitem{BD} N.R.F. Braga and A. Das , Nucl. Phys. B442 (1995) 655.
\bibitem{AB} E.M.C.Abreu and N.R.F.Braga," A superspace Formulation 
for the Master Equation" , to appear in Phys. Rev. D.
\bibitem{FS} L. D. Faddeev, Phys. Lett. B145
(1984) 81; L. D. Faddeev and S. L. Shatashvili, Phys. Lett. B167
(1986) 225.
\bibitem{BM} N. R. F. Braga and H. Montani, Phys. Lett. B264 (1991) 125.
\bibitem{Harada} K. Harada and I. Tsutsui, Phys. Lett. B183 (1987) 311.
\bibitem{SUP} S. Ferrara, O. Piguet and M. Schweda, Nucl. Phys. 
 B119 (1977) 493;  K. Fujikawa, Progr. Theor. Phys.  59 (1978) 2045.
\bibitem{COL} J. Alfaro and P. H. Damgaard, Nucl. Phys.  B404
 (1993) 751.
\bibitem{DJ}F.De Jonghe, The Batalin-Vilkovisky Lagrangian Quantization
scheme with applications to the study of anomalies in gauge theories,
PH.D. thesis K.U. Leuven, HEP-TH 9403143.
\bibitem{BM2} N.R.F.Braga and H.Montani, Int. J. Mod. Phys. A8 (1993)
2569.
\bibitem{GP1} J. Gomis, J. Paris, Nucl. Phys. B395 (1993) 288.
\end{thebibliography}
\end{document}